# Highly sensitive AuNCs@GSH/Ch-PtNPs metal nanoprobes for fluorescent and colorimetric dual-mode detection of ascorbic acid in drink

Wei Zheng·Shuyu Feng·Yanwei Chen·Guiye Shan

**Abstract**
Fluorescence detection is a commonly used analytical method with the advantages of fast response, good selectivity and low destructiveness. However, fluorescence detection, a single-mode detection method, has some limitations, such as background interference that affects the accuracy of the fluorescence signal, lack of visualization of the detection results, and low sensitivity for detecting low-concentration samples. In order to overcome the shortcomings of fluorescence single-mode detection, we used the dual-mode method of fluorescence and colorimetry to detect ascorbic acid. In this paper, gold nanoclusters (AuNCs) and chitosan-modified platinum nanoprobes (Ch-PtNPs) were selected to construct a metal nanoprobe for fluorescence and colorimetric bimodal detection of ascorbic acid (AA), which was effective.The fluorescence excitation and emission wavelengths of AuNCs were 422 nm and 608 nm, respectively. Due to the oxidase activity of PtNPs, it can catalyze the oxidation of the chromogenic agent 3,3,5,5-tetramethylbenzidine (TMB) to generate oxTMB to produce colorimetric signals. At the same time, oxTMB has an energy overlap region with AuNCs, which can produce the fluorescence resonance energy transfer (FRET) effect to excite the fluorescence of AuNCs. And after the addition of AA, the fluorescence signal was restored while the colorimetric signal of oxTMB disappeared. The dual-mode detection of AA by fluorescence and colorimetry in the probe system enhances the specificity and accuracy of the detection. This bimodal detection method solved the problem of low detection sensitivity in the low concentration range of the analytes to be tested, and was linear in the lower (0-50 μM) and higher (50-350 μM) concentration ranges, respectively, and had a lower detection limit (0.034 μM). This glutathione-based gold cluster assay is characterized by simplicity, rapidity and accuracy, and provides a new way for the quantitative analysis of ascorbic acid. In addition, the method was validated during the determination of AA in beverages, which has the advantages of high sensitivity and fast response time.

## Introduction

Gold nanoclusters (AuNCs) have a wide range of applications and research value in the field of imaging and sensing due to their macroscopic quantum tunneling effect, large Stokes shift and long fluorescence lifetime [1,2]. Compared with traditional materials such as carbon dots (CDs), AuNCs have the advantages of good biocompatibility, size-dependent fluorescence properties and high specific surface area. Since AuNCs show great potential for fluorescence detection, they have become a focused sensing and detection material [3-9].

Platinum nanoenzymes (PtNPs) have excellent catalytic activity that mimics the activity of natural enzymes, such as oxidases [10,11]. This catalytic activity enables precious metal nanoenzymes to efficiently catalyze the oxidation reaction of the color developer in colorimetric detection, thus producing visible color changes and achieving the detection of the target substance [12]. PtNPs, as one kind of precious metal nanoenzymes, are widely used in colorimetric detection with the advantages of easy surface modification and good stability [13,14].

Ascorbic acid (AA) is one of the nutrients that the human body needs to ingest, with antioxidant [15], immunomodulation [16], collagen synthesis [17], and promotion of iron absorption [18] of many functions, which plays an important role in maintaining health and preventing diseases [19]. And the human body does not

✉ Yanwei Chen
  yanweichen@nenu.edu.cn

1 Centre for Advanced Optoelectronic Functional Materials Research, Key Laboratory for UV Light-Emitting Materials and Technology of the Ministry of Education, Northeast Normal University, Changchun 130024, China produce AA by itself and needs additional intake. Once the human body ingests insufficient amounts of AA, by itself and needs additional intake. Once the human body ingests insufficient amounts of AA, it can cause problems such as scurvy [20] as well as anemia [21]. Therefore, detecting the amount of ingested AA is of great importance. Traditional methods for detecting AA include electrochemical method [22], photothermal imaging method [23], colorimetric method [24] and fluorescence method [25], etc. Among them, fluorescence detection method has the advantages of high detection sensitivity, low detection limit, simple method, easy to operate and low cost. However, such as the fluorescence method, this single-mode detection limits its application due to factors such as low detection accuracy and susceptibility to interference [26]. Therefore, it is necessary to improve the detection method to increase the detection accuracy. Dual-mode detection makes up for the shortcomings of single-mode, breaks through the limitations of the current single-mode detection of substances, and provides us with a new idea of detection. One of the new methods of fluorescence and colorimetric dual-mode detection of AA can be intuitively and accurately carried out [27].

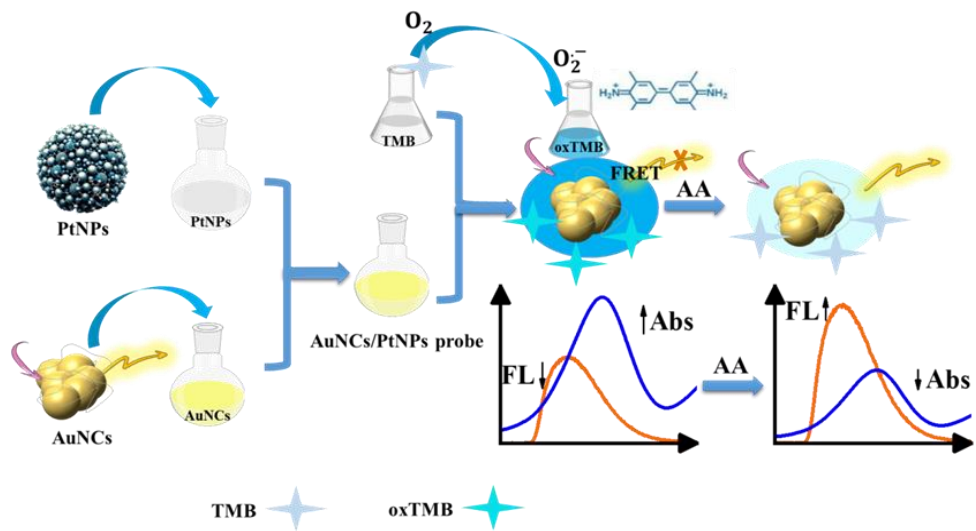

**Scheme 1** Preparation of AuNCs@GSH/Ch-PtNPs probe and principle diagram of double-mode detection of AA sensor by colorimetric fluorescence method.

In this paper, a new idea of dual-mode detection of AA by fluorescence and colorimetric method was pioneered. Scheme 1 is the detection schematic diagram of the detection system. AuNCs that emit fluorescence at 608 nm under 422 nm excitation were synthesized by a one-pot method, based on which PtNPs were added to construct a metal nanoprobe system. The nanoprobes were added into the color developer TMB, and the PtNPs catalyzed the oxidation of TMB to produce oxTMB, which appeared as a colorimetric signal at 652 nm in the UV-vis absorption spectrum, and the solution gradually turned blue. Due to the fluorescence resonance energy transfer (FRET) phenomenon between oxTMB and AuNCs, AuNCs fluorescence was burst by oxTMB. After AA was added, AA provided electrons to oxTMB to reduce it to TMB, and the characteristic peak of oxTMB at 652 nm gradually disappeared, and the solution gradually changed from blue to colorless. With the increase of AA concentration，the fluorescence burst induced by oxTMB was attenuated, and the fluorescence intensity of AuNCs at 608 nm increased. The two-mode detection process of the probe system produced "off-on" fluorescence signal and "on-off" colorimetric signal changes. The fluorescence and colorimetric dual-mode detection of AA by the probe system has the advantages of high sensitivity and fast response in the detection of real samples, which indicates that it has a broad application prospect.

# Experimental section

## Materials

Glutathione (GSH), ascorbic acid ($C_6H_8O_6$), 3,3',5,5'-tetramethylbenzidine ($C_{16}H_{20}N_2$), glucose (Glu), sodium chloride (NaCl), potassium chloride (KCl), L-arginine (Arg), L-cysteine (Cys), L-lysine (Lys) were purchased from Ron Reagent Co. (Shanghai, China).30% hydrogen peroxide ($H_2O_2$), chloroplatinic acid ($H_2PtCl_6$), chitosan (Ch), and sodium borohydride ($NaBH_4$) were purchased from Sinopharm Chemical Reagent Co. Bovine serum albumin (BSA), disodium hydrogen phosphate ($Na_2HPO_4$), and potassium dihydrogen phosphate ($KH_2PO_4$) were purchased from Rinne Technology Development Co Ltd (Shanghai, China). Tetrachloroauric acid trihydrate ($HAuCl_4 \cdot 3H_2O$) was purchased from Aladdin Biochemical Sciencef and Technology Corporation (Shanghai, China).

## Apparatus

Transmission electron microscopy (TEM) images were characterized by a JEM-2100 PLUS TEM (Japan). Energy dispersive X-ray analysis (EDAX) was characterized by Phillips XLZ 30. Absorption spectra were characterized by UV-2600 UV-Vis spectrophotometer. Excitation and emission spectra were tested by PTI-Quanta Master 400 fluorescence spectrometer.X-ray diffraction (XRD) results were characterized by Rigaku D/max-2500 X-ray diffractometer. Fourier transform infrared (FT-IR) spectra were tested by Nicolet 6700-FTIR spectrometer. Free radicals were characterized by an electron paramagnetic resonance spectrometer ruker EMXnano. Particle size was characterized by a particle size and zeta potential analyzer Rigaku Zetasizer Nano S90. pH was tested by a pH meter Mettler-Toledo LE438. Transient fluorescence lifetime spectra were tested by a Spectrum Modular Fluorescence Spectrometer PTI-Quanta Master 8000.

### Preparation of AuNCs and PtNPs nanoprobe systems

According to the reported method, AuNCs@GSH was synthesized using a one-pot hydrothermal method [28], and the detailed preparation scheme is shown in S2. As shown in the flowchart of S3, on the basis of the existing synthesis method [29], we improved to prepare platinum nanoparticles (Ch-PtNPs) using a simple chemical reduction method. After optimization of the conditions tested, 500 μL, 5 mM of TMB solution was added to HAuCl4 solution (500 μL, 0.04 mM), PtNPs solution (900 μL, 0.04 mM), and HAc-NaAc (0.2 mM) buffer solution with pH 4. The mixture was filtered after incubation for one hour and the solution was used for the next step of the experiment.

### Fluorescence and colorimetric dual-mode detection of AA

Various AA solutions with different concentrations were added to the AuNCs and PtNPs nanoprobe systems, and after incubation for a period of time, the absorption and fluorescence spectra of each group of mixtures were recorded. The linear relationship between the degree of colorimetric degradation and fluorescence recovery of the system and the concentration of AA was obtained, and the study of the system's immunity and selectivity to the fluorescence and colorimetric parts of AA detection was carried out.

### Determination of AA distributed in drink

To verify that this experiment for AA detection was also applicable to the detection of AA in drink, various drinks were selected as samples. The drinks were purchased from a local supermarket, and the juices were pretreated using a 0.22 μM filter membrane. A sample of 30 mL of each drink was loaded into a centrifuge tube, sonicated for 20 min, and centrifuged (8000 rpm, 10 min) to obtain the supernatant. The samples were diluted 5-fold, and different concentrations of treated AA were added to the samples, which were analyzed according to the analytical procedure.

## Results and discussion

### Characterization of AuNCs and PtNPs Nanoprobes

The successful synthesis, size and crystal morphology structural characterization of AuNCs@GSH was supported by relevant characterization data. As shown in the transmission electron microscopy TEM images of Fig. 1(a) and (b), the AuNCs@GSH are spherical in shape and dispersed very homogeneously in solution. The lattice planes of the AuNCs are spaced at a spacing of 0.215 nm, which corresponds to the (111) crystallographic plane of the AuNCs, evidencing the presence of the gold nuclei. As shown in Fig. 1(c) the particle size distribution of AuNCs, the size of the gold clusters is around 1-2 nm, and their average size is 1.9 nm, which satisfies the size requirement of AuNCs. The elemental and compositional valence states of AuNCs are characterized. The elemental distribution of AuNCs can be obtained from the EDS in Fig. 1(d), with elements such as Au, S, C and O, which are consistent with AuNCs. The crystal structure of AuNCs was characterized and the XRD images of AuNCs in Fig. 1(e) have diffraction peaks at 38.269°, 44.600° and 64.678°, which correspond to the (111) and (200) and (222) crystal planes, respectively. This corresponds to the lattice crystal plane results of AuNCs in TEM and the Au standard card (PDF#1-0072-Au), proving that AuNCs@GSH was successfully synthesized. Based on the IR spectrogram of AuNCs in Fig. 1(f), compared to GSH, the S-H bond of AuNCs@GSH at 2524 cm-1 disappears and is converted into an Au-S bond immobilized on AuNCs, which aids in verifying the synthesis of AuNCs@GSH.

The optical properties of AuNCs@GSH were characterized. As shown in the UV-vis absorption spectrum of Fig. 2(a), the characteristic absorption edge of AuNCs exists in the range of 200-400 nm, meanwhile the ligand of AuNCs, GSH, does not have an absorption peak at 400 nm. And the apparent absence of plasmon resonance absorption peak at 520 nm for AuNCs is a good indication that there is no formation of large-sized gold nanoparticles and the substance exists in the form of smaller sizes, further validating the synthesis of AuNCs. The inset of Fig. 2(a) demonstrates the color of the AuNCs solution under UV light in room light and in dark environment as yellow and bright orange, respectively. As shown in Fig. 2(b), by de-irradiating the AuNCs solution with different wavelengths of excitation light source (350-440 nm), the positions of the emission peaks of AuNCs corresponding to the monitored wavelengths did not change, and the intensity of the emission peaks was strongest under the excitation of the excitation light source at the wavelength of 420 nm. As shown in the fluorescence excitation and emission spectra of Fig. 2(c), AuNCs have the strongest excitation peak at 422 nm and the strongest emission peak at 608 nm, which is consistent with the

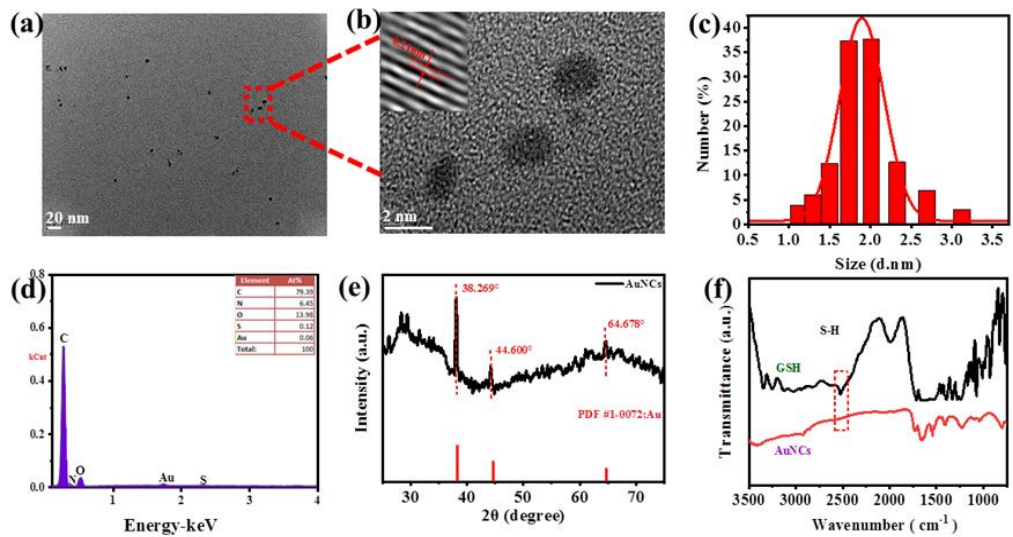

**Fig. 1** (a) Transmission electron microscopy TEM image of AuNCs at 20 nm scale. (b) High-resolution electron microscopy HAADF-STEM image of AuNCs at 2 nm scale. (c) Particle size distribution of AuNCs. (d) EDS energy spectrum of AuNCs. (e) X-ray diffraction (XRD) image of AuNCs and its ligand GSH. (f) Fourier transform infrared (FT-IR) spectra FT-IR images of AuNCs and their ligand GSH

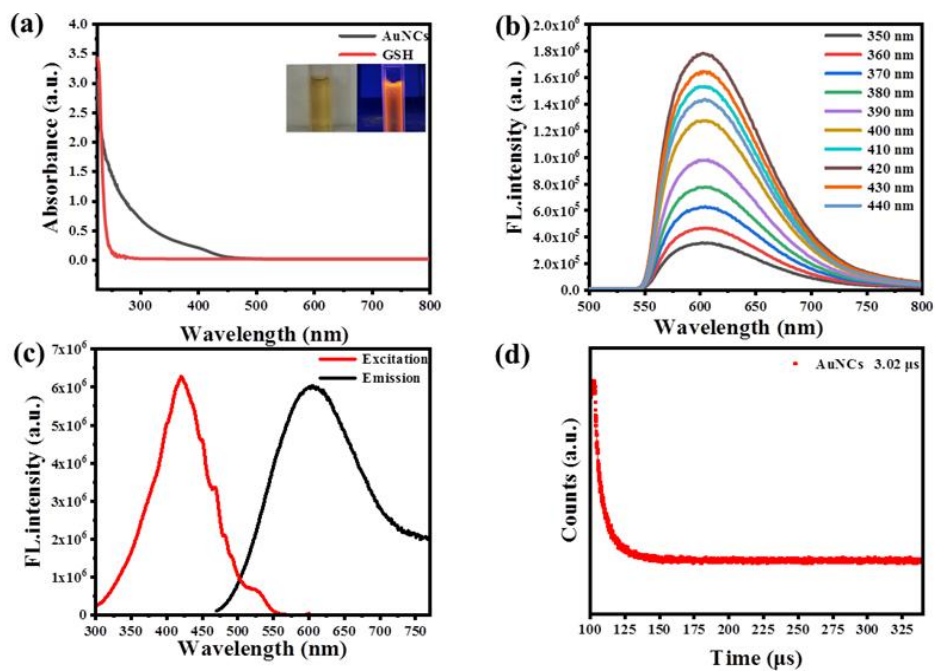

**Fig.2** (a) UV-VIS absorption spectra of AuNCs@GSH and GSH. Illustration: Visible light (left) and ultraviolet light at 365 nm (right). (b) The fluorescence spectra of AuNCs@GSH at different excitation wavelengths in the range of 350-440 nm. (c) the excitation spectrum (red) and emission spectrum (black) of AuNCs@GSH. (d) AuNCs@GSH transient fluorescence lifetime diagram.

conclusion that the emission peak intensity is the strongest under the excitation of the excitation source at 420 nm in Fig. 2(b). Figure 2(d) demonstrates that AuNCs have a long fluorescence lifetime at around 3.02 μs. In addition, the quantum absolute yield of AuNCs@GSH is calculated by using the integral sphere of QM400 to be 3.58%. These demonstrate that AuNCs@GSH has good luminescence performance.

The microstructures of Ch-PtNPs are shown in the TEM images of Fig. 3(a) and the HAADF-STEM images of (b), and the Ch-PtNPs are spherical and dispersed very uniformly in solution. (The inset of (a) shows the columnar distribution of the hydrated particle size of Ch-PtNPs. As shown in the inset particle size distribution graph, the PtNPs have a small size with an average size of 1.75 nm, while according to the previous literature [30,31], the synth-

-esized such Pt nanoparticles are generally around 1-2 nm in size, these TEM images synergistically proved that Pt nanoparticles were successfully synthesized. Their lattice plane spacing is 0.22 nm, which corresponds to the (100) crystal plane of PtNPs. The chemical and surface properties of Ch-PtNPs were characterized by FT-IR spectroscopy. As shown in Fig. 3(c), in the FT-IR spectra of Ch-PtNPs, comparison with the spectra of Ch indicates that the peaks of the N-H bending vibrations at 1645 cm$^{-1}$ and 1600 cm$^{-1}$ are shifted to 1637 cm$^{-1}$ and 1560 cm$^{-1}$, respectively, at the lower wavelengths, the peaks of C-N stretching vibrations at 1325 cm$^{-1}$ and 1260 cm$^{-1}$ almost disappeared, indicating that the peak of $C_3$-O stretching vibrations at 1083 cm$^{-1}$ moved to 1077 cm$^{-1}$. At the same time, as the peaks of $C_1$-O-$C_4$ stretching vibration at 1155 cm$^{-1}$ and $C_6$-O stretching vibration at 1025 cm$^{-1}$ remain These significant changes indicate that the N2 a-

-nd O3 groups of Ch interact with the PtNPs, thus demonstrating that the hydroxyl groups of chitosan and the Pt atoms interacted successfully to form Ch-PtNPs. From the XRD images of the PtNPs in Fig. 3(d), it can be seen that there are diffraction peaks at 39.4°, 45.9°, and 67.3°, which correspond to the (111), (200), and (220) crystal planes. This corresponds exactly to the lattice crystal plane results of the TEM of Ch-PtNPs, again proving the successful synthesis of PtNPs.

In addition, to further explore the composition of elements and surface oxidation valence states in Ch-PtNPs, they were characterized by XPS energy spectroscopy. As shown in Fig. 3(e), the XPS full-spectrum image of Ch-PtNPs consists of elements such as C, N, O, and Pt, which initially proves the successful synthesis of Ch-PtNPs. The peaks located at 532 eV, 401 eV, and 285 eV belong to O1s, N1s, and C1s, respectively. since the main source of catalytic activity of the enzyme comes from the 4f energy level of PtNPs, as shown in Fig. 3(f) 4f energy level of PtNPs. The two peak positions located near 73 eV and 76 eV in the energy spectrum correspond to Pt $4f_{7/2}$ and Pt $4f_{5/2}$, respectively, which are corresponding to the characteristic peaks of Pt. After convolutional calculation, 73.30 eV (purple) and 76.48 eV (blue) correspond to $Pt^{2+}$, and 72.40 eV (green) and 75.80 eV (yellow) correspond to $Pt^0$, which indicates that the surface oxidized valence states of Ch-PtNPs contain $Pt^0$ and $Pt^{2+}$, which further proves that $Pt^{4+}$ is successfully reduced. It is worth noting that the oxidized valence states of $Pt^0$ and $Pt^{2+}$ in Ch-PtNPs are different from those of other common PtNPs, which are generally $Pt^{4+}$ and $Pt^0$ [32].

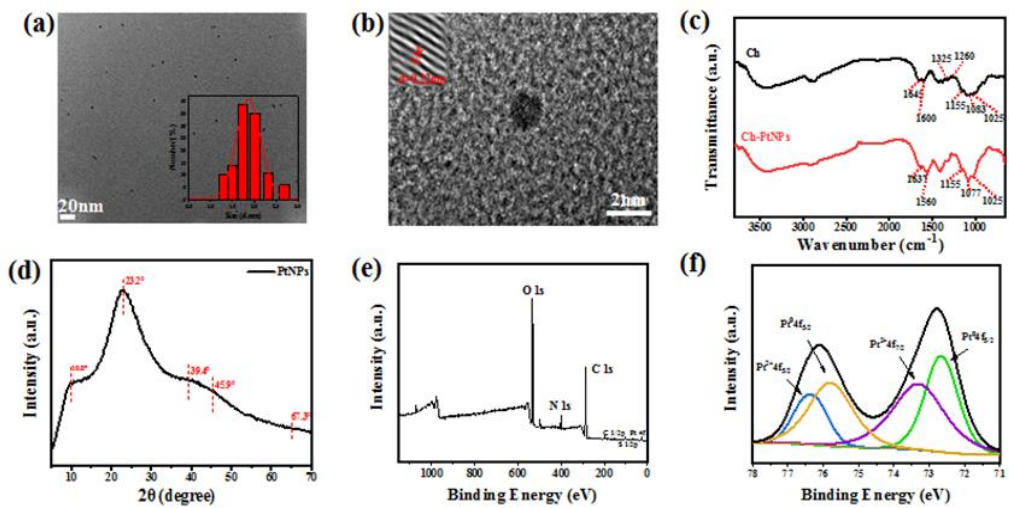

**Fig. 3** (a) TEM image of Ch-PtNPs at 20 nm scale. (b) HAADF-STEM image of Ch-PtNPs at 2 nm scale. (c) FT-IR spectrograms of Ch-PtNPs (red) and Ch (black). (d) XRD spectra of PtNPs. (e) XPS of Ch-PtNPs in the wide-scan measurement energy spectrum. (f) high-resolution energy spectrum of Pt 4f.

The catalytic phenomena and catalytic mechanism of PtNPs were investigated. UV-visible absorption tests were performed on PtNPs, TMB, PtNPs and TMB, and the UV-visible absorption spectra in the range of 200-800 nm wavelength band were observed, and Fig. 4(a) shows the absorption spectra of each of the three. From Fig. a, it can be seen that PtNPs and TMB together, compared with PtNPs, TMB alone, have distinct absorption peaks only at 360 nm and 652 nm. The substance with the characteristic absorption peak at 652 nm is oxTMB. according to the existing literature [33], PtNPs have the property of an oxidase, which catalyzes the oxidation of TMB to form oxTMB, which is a relatively safe color rendering agent with little toxic effect. The oxidase-catalyzed process is that dissolved oxygen in solution is catalytically oxidized by the oxidizing nanoenzyme to generate superoxide radicals, and then the superoxide radicals catalyze the oxidation of TMB to oxTMB. it is known that the temperature and pH affect the enzyme activity, and the optimal conditions for the activity of the PtNPs (temperature and pH) were explored. Figure 4(b) shows the histograms of the distribution of peak absorption of o--xTMB at 652 nm with temperature at different temperatures (20 °C, 25 °C, 30 °C, 35 °C, 40 °C, 45 °C), and the peak absorption of oxTMB in solution was measured after 15 min of incubation. It can be seen that the catalytic activity of the enzyme is the strongest in the range of 20-45 °C, close to the room temperature of 35 °C. From this, it was concluded experimentally that PtNPs had the best enzyme activity at pH 4 and temperature 35 °C. To investigate the optimal catalytic pH of PtNPs, we explored the catalytic conditions. Figure 4(c) shows the bar graph of peak absorption of oxTMB at 652 nm as a function of pH. The UV-Vis absorption test of PtNPs revealed that the catalytic activity of PtNPs for TMB was optimal at pH 4 in the range of pH 1-11, thus determining that the optimal pH for PtNPs is 4. And in order to validate the mechanism of the catalytic oxidation of TMB by PtNPs, we investigated the catalytic conditions by varying the different oxygen contents in the solution (5%, 10%, 15%, 20%, 25%) to observe the UV-Vis absorption spectrum at 652 nm. As shown in Fig. 4(d), the UV-Vis absorption spectra at different dissolved oxygen contents in the solution, with the gradual increase of dissolved oxygen content, the absorption peak has a maximum value at 20%. This side proves that the dissolved oxygen content in the solution has a certain effect on the catalytic oxidation of TMB by PtNPs to generate oxTMB, which aids in verifying the previous mechanism of PtNPs catalytic oxidation of dissolved oxygen in solution to generate superoxide radicals.

In order to have a deeper understanding of the catalytic activity of PtNPs, we have investigated the steady state kinetic properties of PtNPs. The concentration of PtNPs was fixed at 0.4 mM by the controlled variable method, and the concentration of the color developer TMB was changed, and the UV-Vis absorption spectra of oxTMB at 652 nm were recorded at the same time, so as to investigate the change rule of reaction rate with concentration. According to the Lambert-Beer law $A = \varepsilon bc$ can get the enzymatic reaction rate, A is the absorbance of UV-visible absorption, ε is the molar extinction constant ($\varepsilon_{oxTMB}$ is usually 39,000 $M^{-1}\cdot cm^{-1}$), b is the thickness of the cuvette (i.e., the thickness of the absorber), and c is the concentration of the substrate, the rate of the reaction can be calculated by the following formula:

$$v = \frac{A}{\varepsilon b \Delta t}$$

Based on the relationship between the reaction rate and the concentration of the color developer, a curve for the reaction rate and the concentration of TMB was made as in Fig. 4(e), and a linear curve of enzymatic kinetics was fitted as shown in Fig. 4(f), which coincides with a typical Miemann-Menten curve. The Mie-Menten kinetic equation is known to be:

$$V = \frac{V_{max}[S]}{K_m+[S]}$$

$$\frac{1}{V} = \left(\frac{K_m}{V_{max}}\right)\left(\frac{1}{[S]}\right) + \frac{1}{V_{max}}$$

Where $V$ denotes the reaction rate, $K_m$ denotes the Michaelis constant, which is a coefficient indicating the affinity between the enzyme and the substrate, $S$ denotes the concentration of the substrate (i.e., the concentration of TMB in this experiment), and $V_{max}$ denotes the maximum rate of reaction in the reaction, which is the maximum rate of the reaction of the system under the condition of saturating the concentration of the substrate. The smaller the value of Michaelis constant $K_m$, the stronger the affinity between the substrate and the enzyme, and the larger the value of maximum reaction rate $V_{max}$ in the reaction, the stronger the catalytic activity of the enzyme. After double inverse processing of the data in Figure 4(e), a scatter plot image was made with the inverse of the reaction rate as the vertical axis and the inverse of the TMB concentration as the horizontal axis in Figure 4(f), and then the image was fitted linearly. The slope $\frac{K_m}{V_{max}}$ of the straight line was fitted to be 0.08793, and the intercept $\frac{1}{V_{max}}$ of the straight line was 0.07082, which led to the calculation of the $K_m$ value of 1.24 mM and the $V_{max}$ value of 14.12 $Ms^{-1}$, which reveals that PtNPs have well substrate affinity and strong catalytic activity.polymer compound with good hydrophilicity.

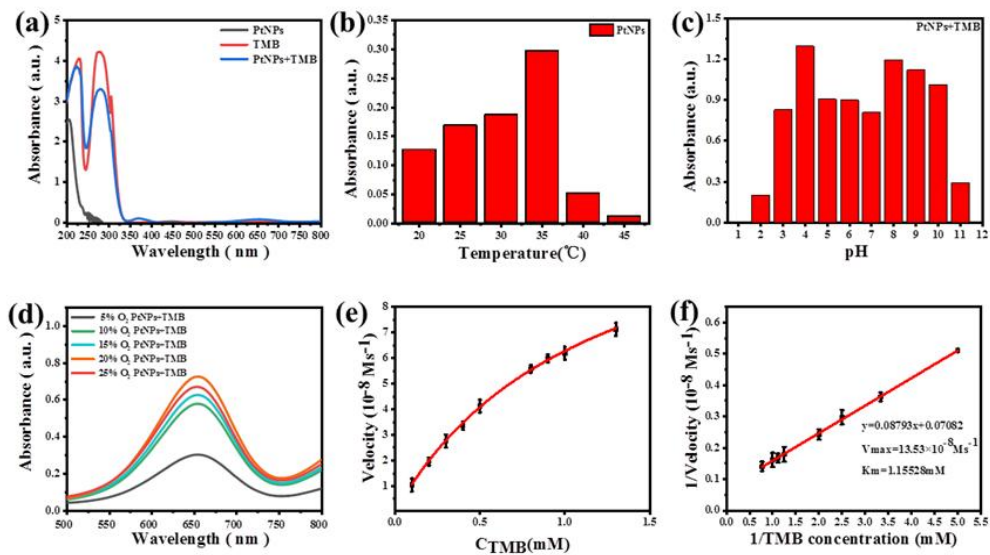

**Fig. 4** **(a)** UV-Vis absorption spectra of PtNPs alone, TMB solution and a mixture of PtNPs and TMB solution all reacted together for 15 min. **(b)** Columnar distribution of absorption values of oxTMB at 652 nm at different temperatures. (c) Histogram of peak absorption values of oxTMB versus pH. **(c)** UV-Vis absorption spectra of the mixed reaction of PtNPs solution and TMB solution at different oxygen contents (5%, 10%, 15%, 20%, 25%) for 15 min. **(d)** Variation curve of colorimetric reaction rate with TMB concentration (fixed PtNPs concentration of 0.4 mM). **(f)** The relationship between the inverse of the colorimetric reaction rate and the inverse of the TMB concentration (the concentration of fixed PtNPs is 0.4 mM).

## Principle of Dual-Mode System Detection

The feasibility of probe fluorescence and colorimetric dual-mode detection of AA was investigated. The probe system dual-mode detection AA is divided into two parts, the colorimetric mode and the fluorescence mode, and the two parts have more obvious signals at the same time. The colorimetric part of the dual-mode detection of AA was realized by UV-Vis absorption spectroscopy. Figure 5(a) shows the UV-visible absorption spectra of AuNCs alone, PtNPs solution, AuNCs and TMB solution mixed, AuNCs and PtNPs solution mixed, PtNPs and TMB solution mixed, AuNCs and PtNPs and TMB solution mixed, and also the UV-visible absorption spectra of AuNCs, PtNPs, TMB, AA all mixed. From Fig. 5(a), the absorption peak at 652 nm is close to zero for the substance alone, and there is no characteristic absorption peak at 652 nm for the addition of TMB or PtNPs to AuNCs respectively, which indicates that AuNCs and TMB, PtNPs do not react with each other. However, when TMB solution was added to PtNPs, there was a more obvious absorption peak at 652 nm relative to other systems, which was due to the excellent catalytic effect of PtNPs catalyzing the generation of superoxide radicals from dissolved oxygen in the solution, which has oxidizing properties and catalytically oxidizes TMB to generate oxTMB, which is possessed by PtNPs. Similarly, there was a more obvious absorption peak at 652 nm when all three of AuNCs, TMB and PtNPs were added, and these phenomena indicated that there was no generation of new complexes in the absorption spectra. These phenomena indicated that no new complexes were generated in the absorption spectra. The absorption value of AuNCs+PtNPs+TMB system decreased after the addition of AA, which provided the feasibility for the colorimetric detection of AA, and achieved the effect of "on-off" of a colorimetric signal.

The feasibility of the fluorescence mode for dual-mode detection of AA by fluorescence colorimetry was realized by observing the fluorescence emission spectra. Figure 5(b) shows the fluorescence emission spectra of AuNCs, PtNPs, TMB, and AA alone as well as AuNCs+PtNPs+TMB, and AuNCs+PtNPs+TMB+AA hybrid systems. The gold nanoclusters AuNCs alone have a more intense fluorescence signal at 608 nm, while the fluorescence is significantly attenuated after PtNPs and TMB, and the fluorescence signal is rapidly recovered after the addition of AA. This process provides the feasibility of the fluorescence method in the bimodal method and leads to an "off-on" change in the fluorescence signal.The specific mechanism of the probe system fluorescence and colorimetric bimodal detection of AA needs to be further explored, for the gold and platinum nanoprobes added to the colorimetric agent TMB such a more obvious fluorescence intensity reduction phenomenon, taking into account that the fluorescence burst of several situations, need to be screened. Figure 5(c) shows the fluorescence emission spectrum of AuNCs and the UV-visible absorption spectrum of oxTMB. From the figure, it can be found that there is a spectral energy overlap between the emission part of AuNCs and the colorimetric mode of oxTMB in the range of 500-800 nm, and theoretically, when the emission spectra of the donor molecules overlap the absorption spectra of the acceptor molecules, a static burst will occur when they are at a close distance ( IFE) or fluorescence resonance energy transfer. The previous UV-Vis absorption spectra showed that no new products were generated during this process, so the static burst in the type of fluorescence burst was excluded.

So there may be a fluorescence phenomenon based on fluorescence resonance energy transfer FRET between AuNCs and oxTMB thus bursting the gold nanoclusters. FRET is an interpretation of the measurement results rather than a measurement technique [34,35]. The condition for FRET to occur is that when the emission spectrum of the donor molecule overlaps with the absorption spectrum of the acceptor molecule, and the spatial distance between the two is close enough, they undergo dipole-dipole interactions, and the fluorescence energy is transferred from the donor molecule to the acceptor molecule, i.e., fluorescence emitted from the acceptor molecule can be observed when the donor molecule is excited at the excitation wavelength of the donor molecule. The spacing of the molecules at 50% energy transfer is called the Förster distance, and for common pairs of FRET molecules, the value of the Förster distance is usually known. In the presence of the acceptor molecule, the change in fluorescence intensity or fluorescence lifetime of the donor molecule can be measured to determine the efficiency of FRET, and thus the distance between the two. The FRET phenomenon in this experiment is that AuNCs act as the energy donor, and oxTMB acts as the energy acceptor, which explains why the fluorescence of the gold clusters in the system of AuNCs+PtNPs+TMB would signal off the FRET can be measured both by fluorescence spectroscopy (intensity) and fluorescence lifetime, and the specific verification of the mechanism principle needs to be analyzed by fluorescence lifetime analysis.

Figure 5(d) shows the transient fluorescence lifetime plots of the mixed systems of AuNCs, AuNCs+PtNPs+TMB, and AuNCs+PtNPs+TMB+AA. Compared with the gold cluster AuNCs alone, the fluorescence lifetime was decreased by adding the color developer TMB to the AuNCs, PtNPs probes, while the fluorescence lifetime was restored by adding AA to the system of AuNCs, PtNPs and TMB, and the process is consistent with the principle of dynamic bursting. So the principle of bimodal detection is that platinum nanoparticles catalyze the oxidation of TMB to form oxTMB, and there is an energy overlap between oxTMB and AuNCs, and FRET phenomenon occurs, at which time the colorimetric signal is turned on and the fluorescence signal is turned off. And after adding AA, oxTMB was reduced, then the FRET phenomenon disappeared, i.e., the colorimetric signal was turned off and the fluorescence signal was turned on.

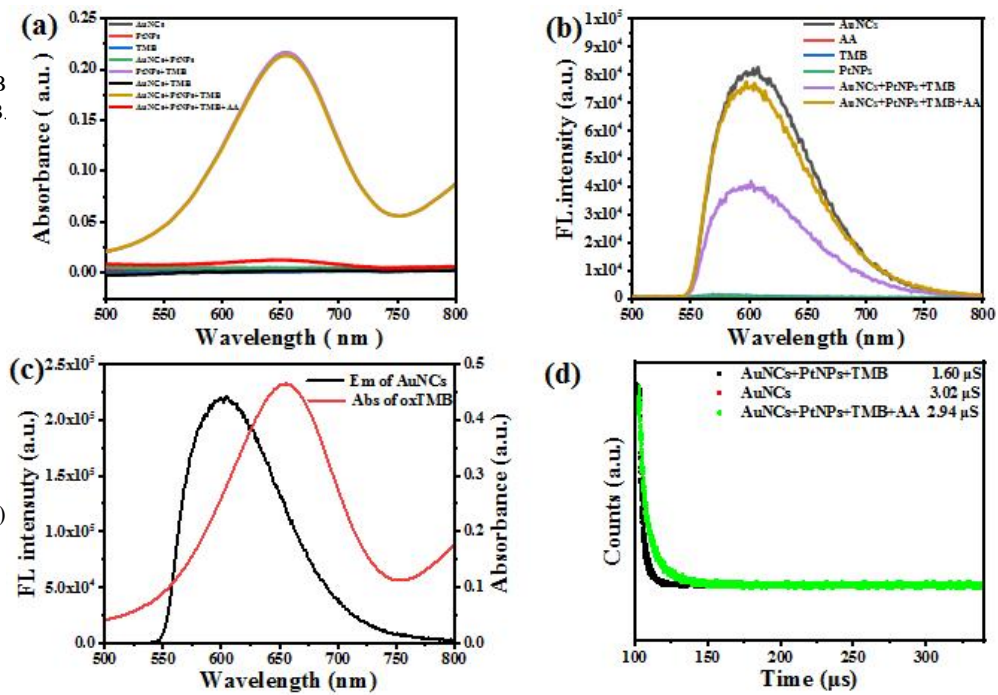

**Fig. 5** (a) UV-Vis absorption spectra of AuNCs, PtNPs, TMB and the systems AuNCs+PtNPs, PtNPs+TMB AuNCs+TMB, AuNCs+PtNPs+TMB. (b) UV-visible absorption spectra of the individual substances AuNCs, PtNPs, TMB, AA and the systems AuNCs+PtNPs+TMB, AuNCs+PtNPs+TMB+AA. (c) Fluorescence emission spectra of AuNCs@GSH and UV-visible absorption spectra of oxTMB (red line is the UV-visible absorption spectrum of oxTMB and black line is the fluorescence emission spectrum of AuNCs@GSH). (d) Comparison of transient fluorescence lifetimes of AuNCs (red) AuNCs+PtNPs+TMB (black) and AuNCs+PtNPs+TMB+AA (green).

Theoretically the fluorescence intensity of AuNCs decreases as the amount of oxTMB increases. In order to further verify the sensing principle of dual-mode detection, we took the amount of dissolved oxygen in the solution (5%, 10%, 15%, 20%, 25%) as the independent variable, and the change of the absorption value of the system and the change of the fluorescence value as the dependent variable, and made the fluorescence colorimetric dual-mode spectra. Figure 6(a) shows the UV-Vis absorption spectra of AuNCs, PtNPs+TMB and AuNCs+PtNPs+TMB at 652 nm for oxTMB at different dissolved oxygen levels, and Fig. 6(b) is the bar chart of the peak of the absorption spectra at 608 nm corresponding to (a) with the change of oxygen content. From the figure, it can be found that AuNCs alone do not undergo a color development reaction and the content of dissolved oxygen has no effect on them. While both PtNPs and TMB were added or all three of AuNCs, TMB and PtNPs were added, the absorption value, i.e., the catalytic effect, was significantly enhanced with the gradual increase of the oxygen content, in which the absorption was saturated when the oxygen content reached 20%, which indicated that the enzyme had the best catalytic effect at this time, and the largest amount of oxTMB was produced.

The fluorescence emission spectra of AuNCs, PtNPs+TMB and AuNCs+PtNPs+TMB at different oxygen contents are shown in Fig. 6(c), and the corresponding fluorescence values at 608 nm of AuNCs, PtNPs+TMB and AuNCs+PtNPs+TMB at 608 nm are shown in the bar graphs with the change of oxygen content in Fig. 6(d). From these two plots, it can be visualized that for AuNCs and AuNCs+TMB alone, the dissolved oxygen content has no effect on the fluorescence intensity of AuNCs at 608 nm. While for the AuNCs+PtNPs+TMB system, it has the best fluorescence burst effect when the oxygen content in solution is 20%. This phenomenon indicates that the solution reaches dissolved oxygen saturation when the amount of dissolved oxygen is 20%, and the content of oxTMB produced reaches the maximum, the fluorescence resonance energy transfer is most obvious, and the best effect is achieved on the bursting of gold clusters of AuNCs, and the result is in line with the previous conjecture, which verifies our conjecture.

**Fig. 6** UV-Vis absorption spectra of **(a)** AuNCs, PtNPs+TMB, and AuNCs+PtNPs+TMB systems at different oxygen contents. **(b)** Histogram of absorption values at 652 nm with oxygen content for AuNCs, PtNPs+TMB and AuNCs+PtNPs+TMB. **(c)** Fluorescence emission spectra of AuNCs, PtNPs+TMB and AuNCs. **(d)** Histogram of fluorescence intensity at 608 nm of AuNCs, PtNPs+TMB and AuNCs+PtNPs+TMB as a function of oxygen content.

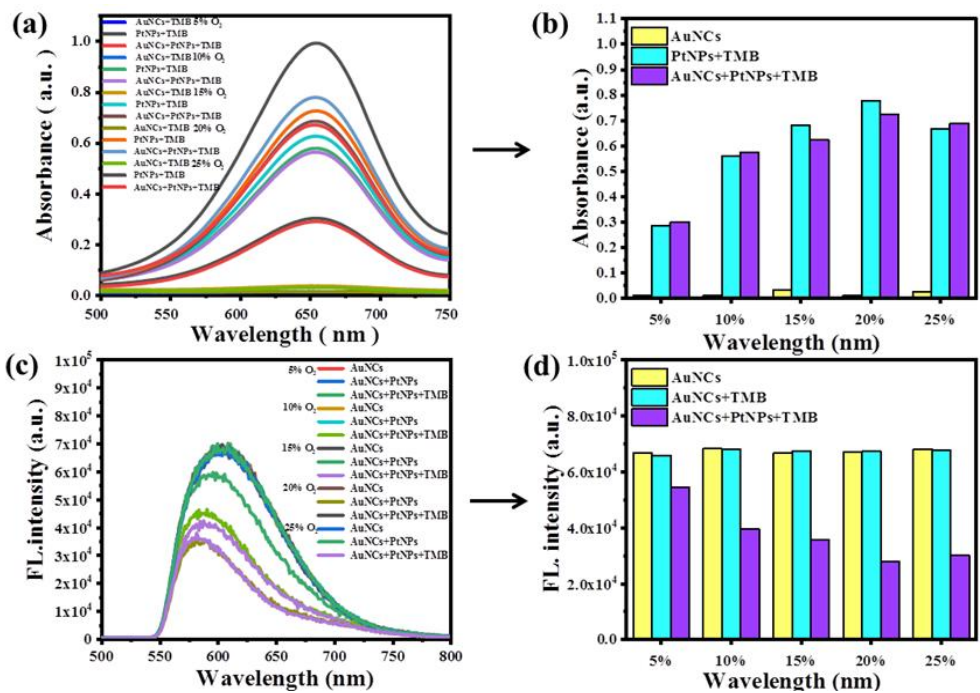

## Detection of ascorbic acid

The optical properties of the probe systems of AuNCs and PtNPs were experimented and analyzed. Figure 7(a) shows the fluorescence emission spectra of AuNCs of the probe system under the excitation of different wavelength light sources. The wavelength of the excitation light source was changed every 10 nm from 350 nm to 440 nm, and it was found that the intensity of the emitted light was enhanced and then weakened with the increase of the wavelength of the excitation light source, and Fig. (a) has the maximum fluorescence emission intensity at 420 nm. Fig. 7(b) is a dot line plot of the colorimetric pattern of AA detection by AuNCs and PtNPs probe system at 652 nm, i.e., the characteristic absorption peak of oxTMB, over time. From the figure, it can be seen that the reaction time of the system to detect AA is around 2 min, and the absorption intensity tends to level off after 2 min, so that a suitable reaction time can be fixed in the subsequent experiments. Figure 7(c) is a line graph of the fluorescence peak of AuNCs at 608 nm versus time for the AuNCs and PtNPs probe system. The graph shows how the fluorescence intensity values of AuNCs and PtNPs probe systems change over 10-60 days, indicating that the AuNCs and PtNPs probe systems can be stored for a long time and are more stable.

**Fig. 7 (a)** Fluorescence emission spectra of AuNCs in the gold/platinum probe system under the excitation of different wavelength light sources. **(b)** Dot line plot of the characteristic absorption peak of oxTMB at 652 nm in the colorimetric mode of AuNCs and PtNPs probe system with the addition of AA to be detected as a function of time. **(c)** Line plot of the peak fluorescence emission at 608 nm in the fluorescence mode of the probe system for AuNCs as a function of time.

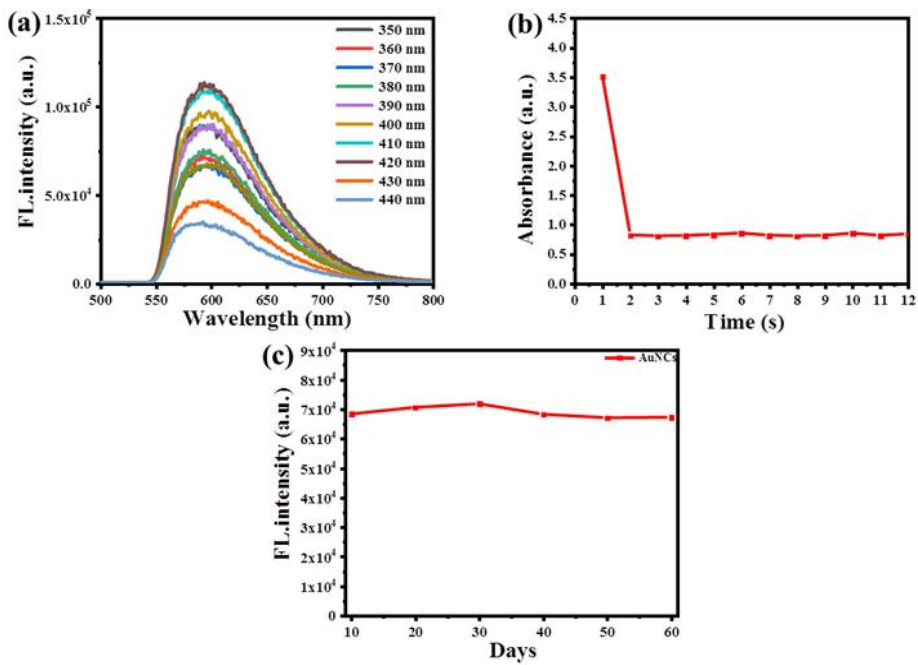

For the fluorescence colorimetric bimodal detection of AA for AuNCs and PtNPs probe systems was investigated. The effect of AA at different concentrations on the fluorescence intensity and absorption intensity of the AuNCs and PtNPs probe system was first explored. Theoretically, as the concentration of AA changes from small to large, the fluorescence signal of the system will gradually turn on and the colorimetric signal will gradually turn off. The reason for this phenomenon is that the AuNCs and PtNPs probe system and the chromogenic agent TMB constitute a system in which the platinum nano-enzymes catalyze the oxidation of TMB to form oxTMB, and there is an energy overlap between oxTMB and the gold clusters in the spectra, and a fluorescence resonance energy transfer occurs, which leads to the fluorescence of the gold clusters being turned off, and the fluorescence signal is restored after the addition of AA accompanied by the turning off of colorimetric signals. Fig. 8(a) and (c) illustrate the occurrence of this phenomenon from the spectral data and changes in the physical photographs. As the concentration of AA increased from 0 μM to 400 μM, the fluorescence intensity of the probe syste tem in the figure gradually increased, accompanied by a gradual decrease in the absorption intensity of the solution. In the insets of (a) and (c), the fluorescence gradually brightens from right to left, while the color of the solution is converted from dark blue to light blue. In addition to analyzing qualitatively, we have to achieve quantitative detection of AA. Figures 8(b) and (d) show the linear plots of peak fluorescence signals and peak absorption in colorimetric mode with AA concentration for AuNCs and PtNPs probe systems. Linear fitting from the 8(b) plot was able to obtain that the fluorescence intensity of AA detected by the probes in the range of 0-350 μM was bisectively linear with the concentration of AA. The linear equation is $Y_1=48.48351X_1+12296.38638$ ($R_1=0.98097$) for AA in the low concentration range of 0-50 μM. And at high concentration 50-350 μM, the linear equation was $Y_2=5.94158X_2+14482.0022$ ($R_2=0.97705$). Similarly, the intensity of the characteristic absorption peaks in the colorimetric pattern of the probe detection of AA also showed a two-band linear relationship with the concentration of AA, as shown in Fig. 8(d). The linear equation is $Y_3=-0.00681X_1+2.00899$ ($R_3=0.96643$) when AA is in the low concentration range 0-50 μM. And the linear equation was $Y_4=-0.00107X_2+1.61934$ ($R_4=0.98215$) for high concentration 50-350 μM. The limit of detection (LOD) for the bimodal assay of gold/platinum probe system was 0.034 μM (S/N=3). As shown in Table 1, compared with previous AA detection methods, the AuNCs and PtNPs probe system has the advantages of high detection sensitivity and wide detection range.

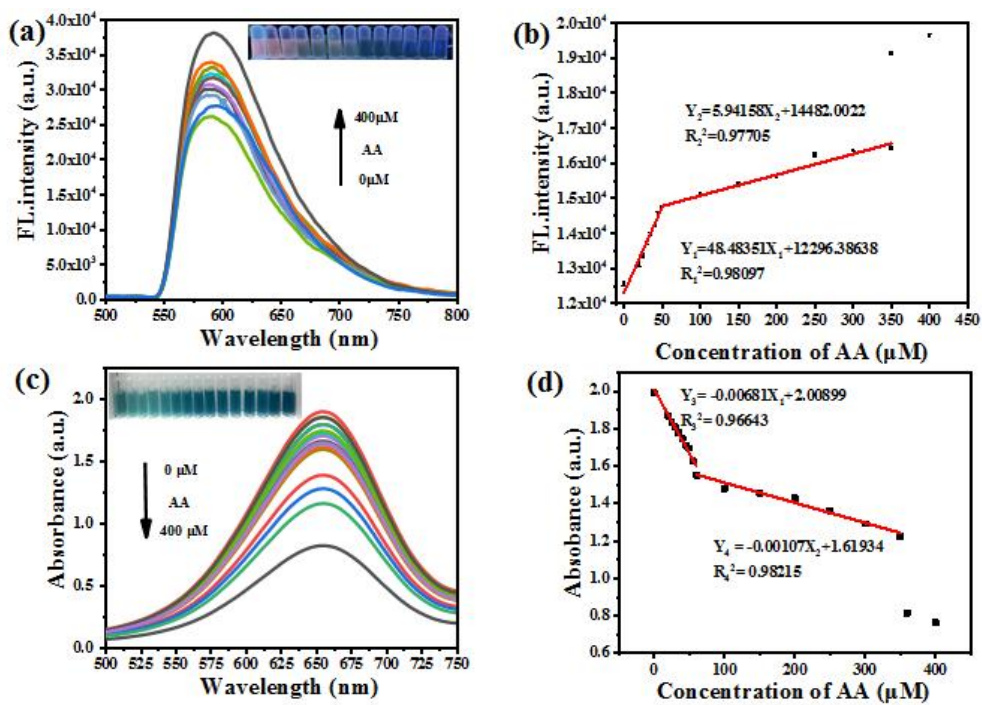

**Fig. 8** **(a)** Effect of different concentrations of AA on the fluorescence intensity of the gold/platinum probe system. The inset shows that the fluorescence color from right to left is gradually brightened with the increase of AA concentration. **(b)** Linear relationship between the peak fluorescence emission at 608 nm of AuNCs in the fluorescence mode of Fig. a and the concentration of AA. **(c)** Effect of different concentrations of AA on the absorption intensity of the Au/Pt probe system. The inset shows the gradual fading of the solution color with increasing AA concentration from right to left. **(d)** Linear relationship between the characteristic absorption peak of oxTMB at 652 nm and AA concentration in the colorimetric mode of Figure c

**Table 1** Comparison table of AA determination by various biosensors

| Nanoparticles used | Method of detection | linear range (μM) | LOD (μM) | refer |
|---|---|---|---|---|
| Au/MnO2@BSA | Fluorescence and magnetic resonance | 4-600 | 0.6 | [36] |
| CNCs | Fluorescence | 0.5-100 | 0.15 | [37] |
| MS GNRs | colorimetric | 0.1-0.25 | 0.049 | [38] |
| MnO$_2$ | colorimetric | 0.25-30 | 0.06 | [39] |
| CuMnO$_2$ NFs | colorimetric | 1-105 | 0.39 | [40] |
| Probe for AuNCs and PtNPs | Fluorescence and colorimetric | 0-350 | 0.034 | this work |

In order to achieve the purpose of better application of the AuNCs and PtNPs probe system for dual-mode detection of AA content in life, we performed anti-interference analysis and selectivity analysis. Figures 9(a) and (c) show the selectivity analysis of colorimetric and fluorescence modes, respectively, (where $\Delta A=A-A_0$, $\Delta F=F-F_0$, A is the absorption value of the system without AA, $A_0$ the absorption value of the system when AA is added, F is the fluorescence intensity of the system when AA is added, $F_0$ is the fluorescence intensity of the system when AA is not added), and from the two figures, we obtained that AuNCs and PtNPs probe systems have a better selectivity for AA with better selectivity. Similarly, in Fig. 9(b) and (d), the dual-mode detection of AA by AuNCs and PtNPs probe systems also has strong immunity to interference. In order to explore the feasibility of this method in drinks, we carried out experiments for the detection of AA in drinks. Different concentrations of AA were added to the drink samples, diluted 1000 times, mixed with the probe system solution and treated for 15 min, and the fluorescence emission spectra and UV-Vis absorption spectra were recorded. All these data were based on three replicate experiments, and the final concentrations of AA in the diluted samples are shown in Table 2. The experimental results were consistent with those shown in the ingredient list of the drinks, indicating that the dual-mode detection system has good prospects for practical application.

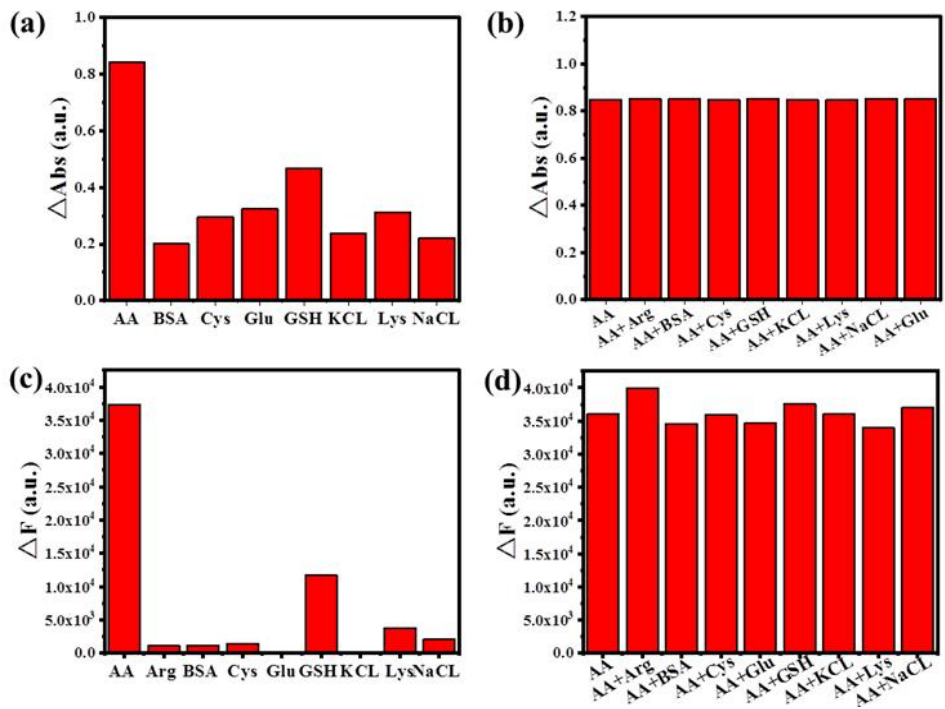

**Fig. 9** (a) Selectivity of the colorimetric mode of the gold/platinum probe system for AA. (b) Anti-interference of AA in the colorimetric mode of gold/platinum probe system. (c) Selectivity of the fluorescence mode of the gold/platinum probe system for AA. (d) Fluorescence pattern of gold/platinum probe system for AA.

**Table 2** Recovery of AA in negative control experiments and spiked drink samples ($n=3$)

| Sample | Found in sample (mM) | Added value (mM) | Total found (mM) | Recovery (%) | RSD (%) |
|---|---|---|---|---|---|
| Sample 0 | 0 | 5 | 4.8 | 96.0 | 3.6 |
|  |  | 10 | 10.2 | 102.0 | 2.8 |
|  |  | 15 | 14.9 | 99.3 | 3.2 |
|  |  | 20 | 19.5 | 97.5 | 2.7 |
| Sample 1 | 3.97 | 5 | 9.11 | 101.5 | 3.1 |
|  |  | 10 | 13.92 | 99.6 | 2.5 |
|  |  | 15 | 18.79 | 99.1 | 2.2 |
|  |  | 20 | 23.68 | 98.8 | 2.8 |
| Sample 2 | 4.67 | 5 | 9.63 | 99.5 | 3.1 |
|  |  | 10 | 14.54 | 99.1 | 2.6 |
|  |  | 15 | 19.31 | 98.2 | 2.1 |
|  |  | 20 | 24.45 | 99.1 | 2.9 |
| Sample 3 | 4.07 | 5 | 8.78 | 96.8 | 3.1 |
|  |  | 10 | 13.94 | 99.1 | 3.1 |
|  |  | 15 | 18.84 | 98.8 | 3.5 |
|  |  | 20 | 23.95 | 99.5 | 2.8 |

## Conclusion

The results showed that the fluorescence and colorimetric dual-mode quantitative detection of AA by AuNCs and PtNPs probe systems had higher sentitivity and accuracy compared with other single-mode detection of AA. PtNPs with more pronounced catalytic effect were employed to catalyze the oxidation of TMB to produce oxTMB, which in turn produced a significant colorimetric signal change at 652 nm in the UV-Vis absorption spectrum. At the same time FRET effect to burst the fluorescence of AuNCs. When AA is added to the probe system, the fluorescence signal is turned off and the colorimetric signal disappears, thus realizing the dual-mode detection of colorimetric "on-off" and fluorescence "off-on", which is able to monitor the changes of the signals in an intuitive and clear way. The dual signal changes of fluorescence and colorimetry enable continuous detection of AA, widening the range of AA detection compared to other sensing systems.

**Supplementary Information**   The contains supplementary material available at PDF"Electronic Supplementary Material"

**Funding**  This work was supported by the National Natural Science Foundation of China (No. 11774048) and the Project from Key Laboratory for UV-Emitting Materials and Technology of Ministry of Education (No. 130028723).

## Declarations

**Conflict of interest**  The authors declare no competing interests.

# Electronic Supplementary Material

## Highly sensitive AuNCs@GSH/Ch-PtNPs metal nanoprobes for fluorescent and colorimetric dual-mode detection of ascorbic acid in drink

Wei Zheng • Shuyu Feng • Yanwei Chen* • Guiye Shan

Centre for Advanced Optoelectronic Functional Materials Research, Key Laboratory for UV Light-Emitting Materials and Technology of the Ministry of Education, Northeast Normal University, Changchun 130024, China

*Corresponding author. E-mail address: yanweichen@nenu.edu.cn.

**Preparation of gold nanoclusters AuNCs@GSH**

In this experiment, we synthesized gold nanoclusters (AuNCs) using glutathione (GSH) as a ligand. GSH has a dual role-acting both as a ligand for stabilizing gold nanoclusters and as a reducing agent. AuNCs@GSH can be synthesized by hydrothermal method, which is widely used for the synthesis of various nanomaterials due to its simplicity, environmental friendliness and mildness to the synthesis conditions.

The specific experimental steps are shown in Figure S2 below. First, 0.09 mL, 101.57 mM of chloroauric acid ($HAuCl_4$) solution and 0.15 mL, 100 mM of GSH were sequentially added to 4.7 mL of ultrapure water. This ratio was chosen to ensure that the reaction of gold ions with GSH was sufficient to promote the formation of high-purity gold nanoclusters. Subsequently, the mixed solution was placed in a water bath, heated to 70 °C and continued to react for 24 h. This step is crucial in the whole synthesis process. High temperature not only accelerates the reduction process of gold ions, but also promotes the formation and growth of gold nanoclusters. After 24 h of reaction, we obtained bright yellow color AuNCs@GSH solution with glutathione as ligand. To remove unreacted raw materials and by-products, the solution was subsequently filtered through a nylon filter head with a diameter of 220 nm. This step is critical for purifying the synthesized samples to ensure the purity and stability of the nanoclusters.

The filtered samples were cryopreserved at 4 °C to maintain their stability and prevent aggregation. Cryopreservation is important for the long-term stability of nanomaterials, especially for nanomaterials in biomedical applications, where stability directly affects their biocompatibility and bioactivity.

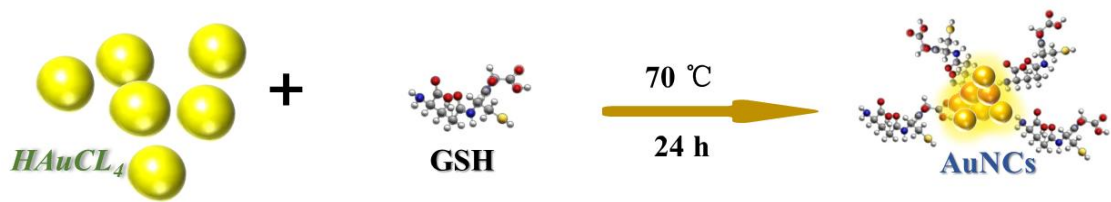

Fig. S2 Synthesis scheme of GSH@AuNCs

**Preparation of platinum nano-enzymes Ch-PtNPs**

As shown in Figure S3, based on the existing synthesis method, we improved the preparation of platinum nanoparticles (Ch-PtNPs) using a simple chemical reduction method. A 0.2% m/v chitosan solution was prepared as a stabilizer using 1% v/v acetic acid solution as a solvent. To this solution, 2 mL, 10 mM/L of $H_2PtCl_6$ solution was added and stirred for 30 min at room temperature to ensure adequate mixing. Slowly add 1 mL, 0.2 M $NaBH_4$ solution, as a reducing agent and continue stirring at room temperature for 90 min. this step completes the reduction process of metallic platinum and the formation of nanoparticles. Three centrifugations were carried out through a hydroalcoholic solvent mixture to purify and collect the product. The product was dried and accurately weighed to obtain the concentration of the product. The resulting solution of Ch-PtNPs needs to be stored in a dark environment at 4 °C to maintain its stability, which can be stored for one year under ideal conditions.

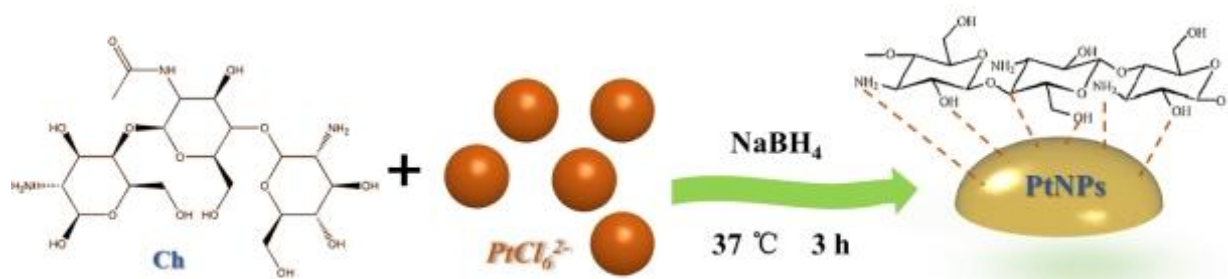

Fig. S3 Synthesis scheme of Ch-PtNPs

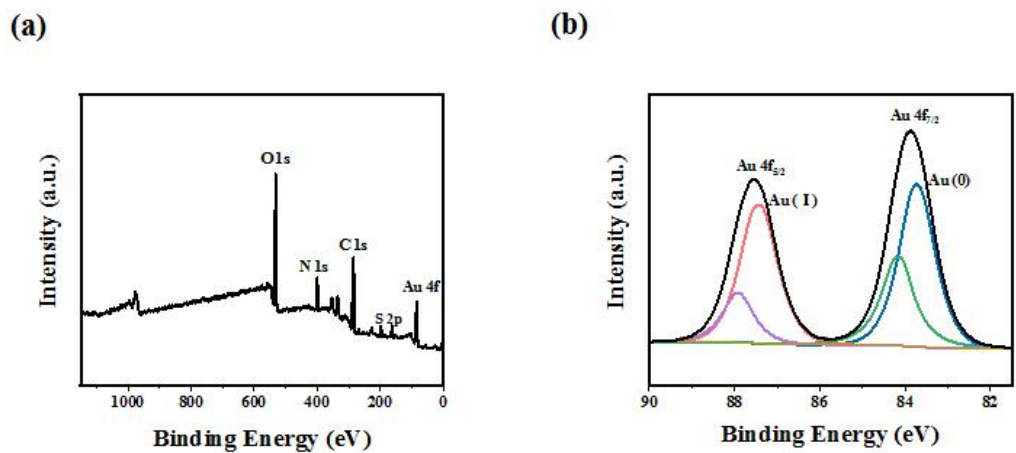

**Figure. S4 (a) XPS energy spectrum of AuNCs-GSH measured by wide scanning. (b) high-resolution XPS energy spectrum of Au 4f**